\documentclass[%
 aip,
 sd,%
 amsmath,amssymb,
 preprint,%
]{revtex4-1}

\usepackage{graphicx}
\usepackage{color}
\usepackage{dcolumn}
\usepackage{bm}

\begin{document}

\long\def \cblu#1#2{{\color{blue}#1 }{\color{red}(#2)}}

\def\FileRef{
\input FName
{
\newcount\hours
\newcount\minutes
\newcount\min
\hours=\time
\divide\hours by 60
\min=\hours
\multiply\min by 60
\minutes=\time
\ 
\advance\minutes by -\min
{\small\rm\em\the\month/\the\day/\the\year\ \the\hours:\the\minutes
\hskip0.125in{\tt\FName}
}
}}

\mathchardef\muchg="321D
\let\na=\nabla
\let\pa=\partial

\let\muchg=\gg

\let\t=\tilde
\let\ga=\alpha
\let\gb=\beta
\let\gc=\chi
\let\gd=\delta
\let\gD=\Delta
\let\ge=\epsilon
\let\gf=\varphi
\let\gg=\gamma
\let\gh=\eta
\let\gj=\phi
\let\gF=\Phi
\let\gk=\kappa
\let\gl=\lambda
\let\gL=\Lambda
\let\gm=\mu
\let\gn=\nu
\let\gp=\pi
\let\gq=\theta
\let\gr=\rho
\let\gs=\sigma
\let\gt=\tau
\let\gw=\omega
\let\gx=\xi
\let\gy=\psi
\let\gY=\Psi
\let\gz=\zeta

\let\lbq=\label
\let\rfq=\ref
\let\na=\nabla
\def\daI{{\dot{I}}}
\def\dsq{{\dot{q}}}
\def\dgj{{\dot{\phi}}}

\def\bgs{\bar{\sigma}}
\def\bgh{\bar{\eta}}
\def\bgg{\bar{\gamma}}
\def\bgF{\bar{\Phi}}
\def\bgY{\bar{\Psi}}

\def\baF{\bar{F}}
\def\bsj{\bar{j}}
\def\baJ{\bar{J}}
\def\bsp{\bar{p}}

\def\hgj{\hat{\phi}}
\def\hgq{\hat{\theta}}

\def\HaT{\hat{T}}
\def\HaR{\hat{R}}
\def\Hsb{\hat{b}}
\def\Hsh{\hat{h}}
\def\Hsz{\hat{z}}

\let\gG=\Gamma
\def\taA{{\tilde{A}}}
\def\taB{{\tilde{B}}}
\def\taG{{\tilde{G}}}
\def\tsp{{\tilde{p}}}
\def\tgF{{\tilde{\Phi}}}

\def\wgx{{\boldmath{\xi}}}

\def\wse{{\bf e}}
\def\wsi{{\bf i}}
\def\wsj{{\bf j}}
\def\wsn{{\bf n}}
\def\wsr{{\bf r}}
\def\wsu{{\bf u}}
\def\wsx{{\bf x}}

\def\vaB{\vec{B}}
\def\vse{\vec{e}}
\def\vsl{\vec{l}}
\def\vgn{\vec{\nu}}
\def\vgk{\vec{\kappa}}
\def\vgt{\vec{\gt}}
\def\vgx{\vec{\xi}}

\def\waA{{\bf A}}
\def\waB{{\bf B}}
\def\waE{{\bf E}}
\def\waJ{{\bf J}}
\def\waV{{\bf V}}
\def\waX{{\bf X}}

\def\R#1#2{\frac{#1}{#2}}
\def\btbl{\begin{tabular}}
\def\etbl{\end{tabular}}
\def\bqbl{\begin{eqnarray}}
\def\eqbl{\end{eqnarray}}
\def\ebox#1{
  \begin{eqnarray}
    #1
\end{eqnarray}}


\preprint{AIP/123-QED}

\title{On the inward drift of runaway electrons during the plateau phase of
    runaway current}  
\author{Di Hu}
 \altaffiliation[While visiting at ]{PPPL, Princeton, New Jersey}
 \email{hudi\_2@pku.edu.cn}
\affiliation{
School of Physics, Peking University, Beijing 100871, China.
}

\author{Hong Qin}
\affiliation{
Princeton Plasma Physics Laboratory, Princeton University, Princeton,
New Jersey, 08540, USA
}
\affiliation{
School of Nuclear Science and Technology and Department of Modern
Physics, University of Science and Technology of China, Hefei, 230026,
China.
}

\date{\today}

\begin{abstract}
  The well observed inward drift of current carrying runaway electrons
  during runaway plateau phase after disruption is studied by
  considering the phase space dynamic of runaways in a large aspect
  ratio toroidal system. We consider the case where the toroidal field
  is unperturbed and the toroidal symmetry of the system is
  preserved. The balance between
  the change in canonical angular momentum and the input of mechanical
  angular momentum in such system requires runaways to drift
  horizontally in configuration space for any given change in momentum
  space. The dynamic of this drift can be obtained by integrating the
  modified Euler-Lagrange equation over one bounce time. It is
  then found that runaway electrons will always drift inward as long as
  they are decelerating. This drift motion is essentially non-linear,
  since the current is carried by runaways themselves, and any runaway
  drift relative to the magnetic axis will cause further displacement of
  the axis itself. A simplified analytical model is constructed to
  describe such inward drift both in ideal wall case and no wall case,
  and the runaway current center displacement as a function of parallel
  momentum variation is obtained. The time scale of such
  displacement is estimated by considering effective radiation drag,
  which shows reasonable agreement with observed displacement time
  scale. This indicates that the phase space dynamic studied here plays
  a major role in the horizontal displacement of runaway electrons
  during plateau phase.

\end{abstract}


\maketitle

\section{Introduction}
\label{s:Intro}

\vskip1em

Large quantity of relativistic runaway electrons is one of the most
feared by-product of tokamak disruption, especially for large devices
with higher total plasma current and higher poloidal magnetic flux
\cite{Rosenbluth1997}. Those highly relativistic electrons are the
direct result of high toroidal inductive field during disruption, which
in turn is the consequence of drastically arising bulk plasma
resistivity as the thermal energy is mostly lost after thermal
quench \cite{Smith2005,Smith2006}. If left unchecked, runaway electrons
can multiply exponentially by Coulomb-collision avalanche
\cite{Rosenbluth1997}, and up to 70\% of initial plasma current can be
converted into relativistic runaway current, forming the so called
``runaway current plateau" \cite{ITERProgress}. Furthermore, the high
energy electrons will keep being accelerated until effective radiation
drag from synchrotron radiation and bremsstrahlung radiation finally
balance the toroidal inductive field
\cite{Martin1998,Andersson2001,Bakhtiari2005}. This will result in a
highly anisotropic relativistic electron beam with energy on the order
of tens of MeVs \cite{Hollmann2013}, as well as a "bump on the tail" kind
of distribution function in the momentum-space
\cite{Aleynikov2015,Decker2015,Hirvijoki2015}. 

The evolution of runaway electrons in momentum-space has been under
substantial investigation during past decades
\cite{Martin1998,Andersson2001,Bakhtiari2005,
  Fussmann1979,Parks1999,Liu2015}. However, the corresponding evolution
in configuration space has not received due attention. During the
aforementioned runaway current plateau, it is widely observed that there
is a gradual inward drift of runaway current
\cite{Gill2002,Eidietis2012,Zhang2012}. This inward drift will
ultimately result in the intersection between runaway electrons and the
wall, causing tremendous damage to the first wall due to its localized
way of energy deposition \cite{Putvinski1997}. The reason of this
displacement is attributed to the force imbalance under
externally generated vertical field \cite{Eidietis2012},
while the possible role played by the dynamic of relativistic electrons
in a self-generated magnetic field has not been fully explored. 

Similar horizontal drift of runaway orbit has been studied using
test particle model \cite{Guan2010}. 
It is found that the balancing of canonical angular momentum budget will
induce a trajectory drift to compensate any change in mechanical angular
momentum, resulting in horizontal motion if runaways are accelerated or
decelerated. This horizontal drift is directional, as opposed to the
diffusion-like behavior of stochastic scattering
\cite{RechRosen1978,Mynick1981}. However, the result of
Ref.\,\onlinecite{Guan2010} can not be directly applied to the
aforementioned inward drift, due to the fact that the current during
plateau phase is carried by runaway electrons themselves.  Thus its
crucial for us to go beyond test particle model and consider the runaway
orbit drift as an nonlinear process, so that any drift relative to the
magnetic axis will result in further displacement of axis itself.

In this paper, the aforementioned inward drift is studied
by considering the current carrying runaway electrons in an 2D
equilibrium during runaway plateau phase. Those runaways are 
being decelerated by effective radiation drag as the original inductive
accelerating field is greatly reduced during plateau \cite{Gill1993}. 
It is found that the runaway current always move inward due to the
balance between canonical angular momentum change and mechanical angular
momentum input if their momentum is decreasing. 
It is also found that the eddy current and the vertical
field are important in stabilizing this inward drift. In the absence
of both, the runaways will not stop until it hit the first wall even for
very small amount of momentum loss. A characteristic time scale is
estimated by considering the synchrotron radiation and bremsstrahlung
radiation drag, and the result is found to reasonably agree with
experimental observations. This agreement indicates the inward drift
motion we discuss here plays an important role in understanding runaway
displacement during plateau phase.

The rest of the paper will be arranged as follows. In Section
\ref{s:AngularMomentum}, the transit orbit of runaway electrons will be
given by seeking its constant canonical angular momentum of runaways. In
Section \ref{s:OrbitDrift}, we consider the displacement 
of runaway current center for any variation of parallel momentum. The
zeroth order drift of runaway current will be given as a function of
runaway momentum change for both ideally conducting wall case and no
wall case. Further, a characteristic time scale of such drift will be
estimated using effective radiation drag. In Section \ref{s:Conclusion},
a conclusion of the work will be given. 

\section{Transit orbit of runaway electrons}
\label{s:AngularMomentum}

\vskip1em

\begin{figure*}
\centering
\noindent
\btbl{c}
\parbox{3.in}{
  \includegraphics[scale=0.7]{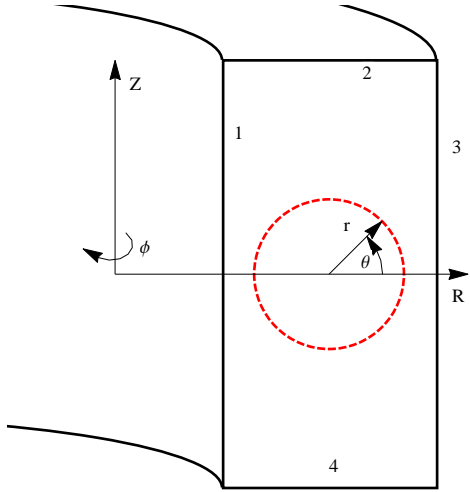}
}
\etbl
\caption{A schematic plot for the cross section of the system of
  interest. The wall is seen as a rectangle toroid as
  shown in the figure by the black solid lines. The red dashed circle
  represent the cross section of runaway torus on this $R Z$ plane. The
  two coordinate system $\left(R,-\gj,Z\right)$ and
  $\left(r,\gq,\gj\right)$ are also shown in the figure.
}
\label{fig:1}
\end{figure*}

We consider a large aspect ratio toroidal system with major radius
$R$, while $R_0$ is defined as major radius corresponding to the
geometry center of the poloidal cross section of the system. For
simplicity, we consider the first wall to be a rectangle toroid
elongated along Z direction. Let the short side of the rectangle be
$2a$, while the long side of it be $4a$. The inverse aspect ratio
$\ge\equiv a/R_0$ is a small number. Four walls of the toroid are
designated by numbers respectively.

A schematic plot of the system of
interest is shown in Fig.\,\ref{fig:1} along with two coordinate systems
$\left(R,-\gj,Z\right)$ and $\left(r,\gq,\gj\right)$. It should be noted
that $R_0$ does not necessarily correspond to the runaway current
center. Since we are primarily interested in the orbit drift of
runaways, no velocity space instabilities will be discussed. Also, since
the vertical stability of the runaway current is essentially a
equilibrium problem which is a separate topic from what we are concerned
here, it will not be treated in our consideration as well.

In the absence of radiation drag, 
we will obtain the transit orbit of runaway electrons 
by seeking its constant canonical angular momentum surface. An easy way
to see how this is done is to realize that the parallel momentum $p_\|$
is a near-constant across the transit orbit for runaway
electrons, as the variation of perpendicular kinetic energy $\gD
\left(\gm B\right)$ is of $\mathcal{O}\left(\ge^3\right)$ comparing to
$p_\|c$ if we assume $p_\bot/p_\|\sim\ge$. Thus the invariance of
canonical angular momentum $p_\gj\left(p_\|,R,-\gj,Z\right)$ defines a 2D
trajectory surface in configuration space for runaway electrons. A more
rigorous consideration would write $p_\|$ as a function of Hamiltonian
$H$ and configuration space coordinates $p_\|\left(H,R,-\gj,Z\right)$,
then we have $p_\gj=p_\gj\left(H,R,-\gj,Z\right)$. The invariance of $H$
and $p_\gj$ in time again defines the trajectory surface \cite{Qin2009}.
It should be noted that, due to the separation of time scale between the
runaway electron's bounce time and their deceleration time, the
trajectory within one bounce period can still be defined by the
near-conservation of canonical angular momentum even when the radiation
drag is included.

In our consideration, all of the runaways are
assumed to be located on a torus with minor radius $a_R$, and with a
single energy and pitch angle. While this is certainly not realistic, it
serves to demonstrate the most fundamental physical idea. In reality,
the runaway electrons have a distribution both in configuration space
and in velocity space, but the well known hollowed image of runaway
radiation strongly suggest a hollowed spatial profile which peaks at
certain minor radius \cite{Zhang2012,Plyusnin2012,Zhou2013}, justifying
our spatial assumption for the runaways as a zeroth order approximation. On
the other hand, the single energy assumption is intended to mimic the
``bump on tail'' distribution of runaways in velocity space, as well as
to greatly simplify the model. The effect of eddy current as a result of
current center motion is taken into account by considering a simplified
ideally conducting wall. This ideally conducting wall will
stabilize current displacement, thus serving as a maximum
stabilization scenario. In real tokamak, it's effect will be
reduced by finite resistivity. 

Since assuming all the runaways are of the same energy and pitch angle,
its sufficient for us to write down the Lagrangian of a
single runaway electron to describe dynamic of the whole runaway
torus. We write down the relativistic guiding center Lagrangian for
runaways in the absence of radiation as follows \cite{Guan2010},
\bqbl
\lbq{eq:Lagrangian}
L\left(\wsx,\dot{\wsx},t\right)
=
\left[
e\left(
\waA_R
+\waA_w
+\waA_{ex}
+\waA_c
\right)
+p_\|\Hsb
\right]
\cdot\dot{\wsx}
-\gg mc^2
.\eqbl
Here, $e$ is the charge of electron, $m$ is electron mass, $c$ is the
speed of light, $\Hsb$ denotes the direction of magnetic field which is
largely in toroidal direction due to the strong toroidal guide
field. $\gg$ is the relativistic factor 
\bqbl
\gg
=
\sqrt{1+\R{p_\|^2}{m^2c^2}+\R{2\gm B}{mc^2}}
.\eqbl
$B$ stand for the magnetic field, and the magnetic momentum is $\gm\equiv
p_\bot^2/2mB$, while $p_\|$ and $p_\bot$ are the momentum parallel and
perpendicular to the field line, respectively. 

We now look at the contribution from vector potentials term by term,
$\waA_R$ is the vector potential generated by the runaway current,
$\waA_w$ is the vector potential corresponding to eddy current generated
in a ideally conducting wall as a reaction to runaway current
motion. Thus $\waA_R+\waA_w$ describe the total vector potential of a
runaway current loop surrounded by the first wall. Apart form those
contributions, $A_{ex}$ corresponds to an additional toroidal electric
field which is generated by external coil and has the following form,
\bqbl
\lbq{eq:ExTorField}
\waE_{ex}\left(R\right)
=
-\R{\pa \waA_{ex}}{\pa t}
,\eqbl
\bqbl
\waE_{ex}
=
E_{ex0}\R{R_0}{R}\hgj
.\eqbl
We should point out that, since we are considering runaway electrons
with high energy, the current carried by those electrons is just
\bqbl
I_R
=
N_Rec
.\eqbl
Here, $N_R$ is the total runaway population. Hence we know that the
kinetic energy change of those electrons will only have minimal impact
on the current itself, so that the inductive electric field from the
change of poloidal magnetic flux is negligible. In a more realistic
consideration, the distribution of runaways in velocity space has to be
considered, and there may be small inductive field exist due to low
energy runaways slowing down thus reducing the runaway current. However,
those inductive field would be much smaller than the toroidal field at
the beginning of current quench due to the much slower current decay
rate. 

Last, there is an additional contribution $\waA_c$ representing
the constant magnetic field imposed by external coils, which include a
toroidal field along $\gj$ direction and a vertical field along $Z$
direction
\bqbl
\waB_c
=
\waB_T
+\waB_Z
,\eqbl
\bqbl
\waB_T
=
-\R{B_{T0}R_0}{R}\hgj
,\quad
\waB_Z
=
B_{Z0}\Hsz
.\eqbl
The vertical field here represents the externally applied position
control field, which will keep the current at the center of the system
at the beginning of our consideration. It serves as a simplified mimic
of the horizontal position control field in a real tokamak, as it is
constant in space and time, as opposed to the real field which varies in
both. Nonetheless, any gradual spatial variation or active position
control can be treated as additional effects, while we are only
concerned with the fundamental trend of runaway drift here.
Due to the form of those constant field, $\waA_c$ can be chosen to have
the following form
\bqbl
\waA_c
=
\R{1}{2}\ln{\left(\R{R}{R_0}\right)}R_0B_{T0}\Hsz
-\R{R_0B_{T0}z}{2R}\HaR
+\R{1}{2}B_{Z0}R\hgj
.\eqbl
Only the $\gj$ component of $\waA_c$ will contribute to the trajectory of
runaway electrons. The constant $B_{Z0}$ is chosen so that at the
beginning of the runaway plateau the runaway current center coincide
with the geometry center of the system $R_0$.

\begin{figure*}
\centering
\noindent
\btbl{c}
\parbox{3.in}{
  \includegraphics[scale=0.3]{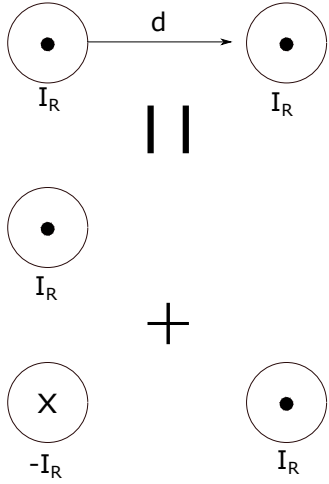}
}
\etbl
\caption{A schematic plot for the treatment of current center
  displacement. The displaced current is effectively represented by
  adding two new current with value $-I_R$ and $I_R$ respectively.
}
\label{fig:2}
\end{figure*}

Now the vector potential contribution from the runaway current itself
will be write down explicitly. We assume \emph{a priori} that the radial
variation of runaway orbit along $\gq$ direction is of
$\mathcal{O}\left(\ge a_R\right)$, so that the poloidal cross-section of
runaway orbit can be approximated as a circle. Hence the magnetic field
directly generated by the runaway current is
axis-symmetric with regard to the runaway current center in the large
aspect ratio limit. We will check the validity of this assumption
\emph{a posteriori}. This yields the following simple contribution 
\bqbl
\waB_\gq
=
\R{\gm_0I_R}{2\gp r}\hgq
,\eqbl
\bqbl
\lbq{eq:RunawayPsi}
\waA_R
=
-\R{\gm_0I_R}{2\gp}\R{R_0}{R}
\left[
\ln{\left|\R{r}{a}\right|}K\left(r-a_R\right)
+\ln{\left|\R{a_R}{a}\right|}I\left(r-a_R\right)
\right]\hgj
,\eqbl
\bqbl
K\left(x\right)
=
1
,\quad \left(x\geq0\right)
;\quad
K\left(x\right)
=
0
,\quad \left(x<0\right)
;\quad
I\left(x\right)
=
1-K\left(x\right)
.\eqbl
Here, $r$ is the minor radius of runaway electrons relative to the
runaway current center. The step function $K$ and $I$ represent the fact
that there is no current within the runaway torus, thus the runaway
current contribution to the poloidal field is zero within the torus, and
the vector potential have a simple $R_0/R$ behavior. Further, the
response from the ideally conducting wall will be treated by simple
magnetic image method. We treat the movement $d$ of runaway current
$I_R$ effectively as adding a pair of new current, one at the original
position of the current and with value $-I_R$ which cancels the original
current, the other at distance $d$ and with value $I_R$ which
represents the moved current. The image currents corresponding to those
two effective currents then represent the eddy current contribution to
current center displacement. A schematic plot of this treatment is
shown in Fig.\,\ref{fig:2}. 

This yields
\bqbl
\lbq{eq:WallPsi}
\waA_w
=
\waA_w^{(+)}
+\waA_w^{(-)}
,\eqbl
\bqbl
\waA_w^{(+)}
=
\R{\gm_0I_R}{2\gp}\R{R_0}{R}
\left(
\ln{\left|\R{r_1^{(+)}}{a}\right|}
+\ln{\left|\R{r_2^{(+)}}{a}\right|}
+\ln{\left|\R{r_3^{(+)}}{a}\right|}
+\ln{\left|\R{r_4^{(+)}}{a}\right|}
\right)
\hgj
,\eqbl
\bqbl
\waA_w^{(-)}
=
-\R{\gm_0I_R}{2\gp}\R{R_0}{R}
\left(
\ln{\left|\R{r_1^{(-)}}{a}\right|}
+\ln{\left|\R{r_2^{(-)}}{a}\right|}
+\ln{\left|\R{r_3^{(-)}}{a}\right|}
+\ln{\left|\R{r_4^{(-)}}{a}\right|}
\right)
\hgj
.\eqbl
Here, $r_i^{(\pm)}$ represents the distance between runaway and the positive
and negative image current centers generated by corresponding wall as
designated in Fig.\,\ref{fig:1} respectively. For leading order
contribution, it would be well enough for us to just take the four pairs of
``primary'' image currents directly corresponds to the current center
displacement. 

Finally, using above equations, Eq.\,(\rfq{eq:Lagrangian}) can be
rewritten as follows,
\bqbl
\lbq{eq:Lagrangian2}
L
=
p_r\dot{r}
+p_\gq\dot{\gq}
+p_\gj\dot{\gj}
-H
,\eqbl
\bqbl
p_r
=
\R{1}{2}e\ln{\left(\R{R}{R_0}\right)}R_0B_0\sin{\gq}
-e\R{R_0B_0z}{2R}\cos{\gq}
,\eqbl
\bqbl
p_\gq
=
\R{1}{2}e\ln{\left(\R{R}{R_0}\right)}R_0B_0r\cos{\gq}
-e\R{R_0B_0rz}{2R}\sin{\gq}
+\left(p+eA_d\right)r\sin{\ga}
,\eqbl
\bqbl
\lbq{eq:AngMomentum}
p_\gj
=
\left[
e\left(A_R+A_w+A_{ex}+\R{1}{2}B_{Z0}R\right)
+p_\|\cos{\ga}
\right]R
,\eqbl
\bqbl
H
=
mc^2\sqrt{1+\R{p_\|^2}{m^2c^2}+\R{2\gm B}{mc^2}}
.\eqbl
Here, $\ga$ is defined as $\tan{\ga}=B_\gq/B_{T}$, so that
$\cos{\ga}\sim 1$ for a large aspect ratio torus, and it can
be approximately seen as a constant. It can be seen from
Eq.\,(\rfq{eq:Lagrangian2}) that there is no explicit dependence on $\gj$
in the Lagrangian, so that the Euler-Lagrange equation yields 
\bqbl
\lbq{eq:Invariant}
\R{\pa L}{\pa \gj}
=
\R{d}{dt}\left(\R{\pa L}{\pa\dgj}\right)
=
\R{d}{dt}p_\gj
=
0
.\eqbl
That is, the symmetry of the system demands the canonical angular
momentum of runaway electron to be a invariant in time. This invariant
will define the surface of runaway orbit in configuration space. 
In the presence of non-conservative forces such as radiation drag, the
toroidal component of the modified Euler-Lagrange equation write
\cite{GoldsteinBook} 
\bqbl
\lbq{eq:ELequationNew}
\R{d}{dt}\left(\R{\pa L}{\pa \dgj}\right)
-\R{\pa L}{\pa \gj}
=
Q_\gj
.\eqbl
Here, $Q_\gj$ corresponds to the change of angular momentum caused by
radiation drag. Thus, for runaway electrons at any given time $t$, we
have
\bqbl
p_\gj\left(t\right)
=
p_\gj\left(0\right)
+\int_0^t{Q_\gj dt}
.\eqbl
Here, $\int_0^t{Q_\gj dt}$ is the total mechanical angular momentum change
caused by radiation drag. It is averaged along the trajectory of runaway
electrons, thus is only the function of time.

\begin{figure*}
\centering
\noindent
\btbl{c}
\parbox{3.in}{
  \includegraphics[scale=0.5]{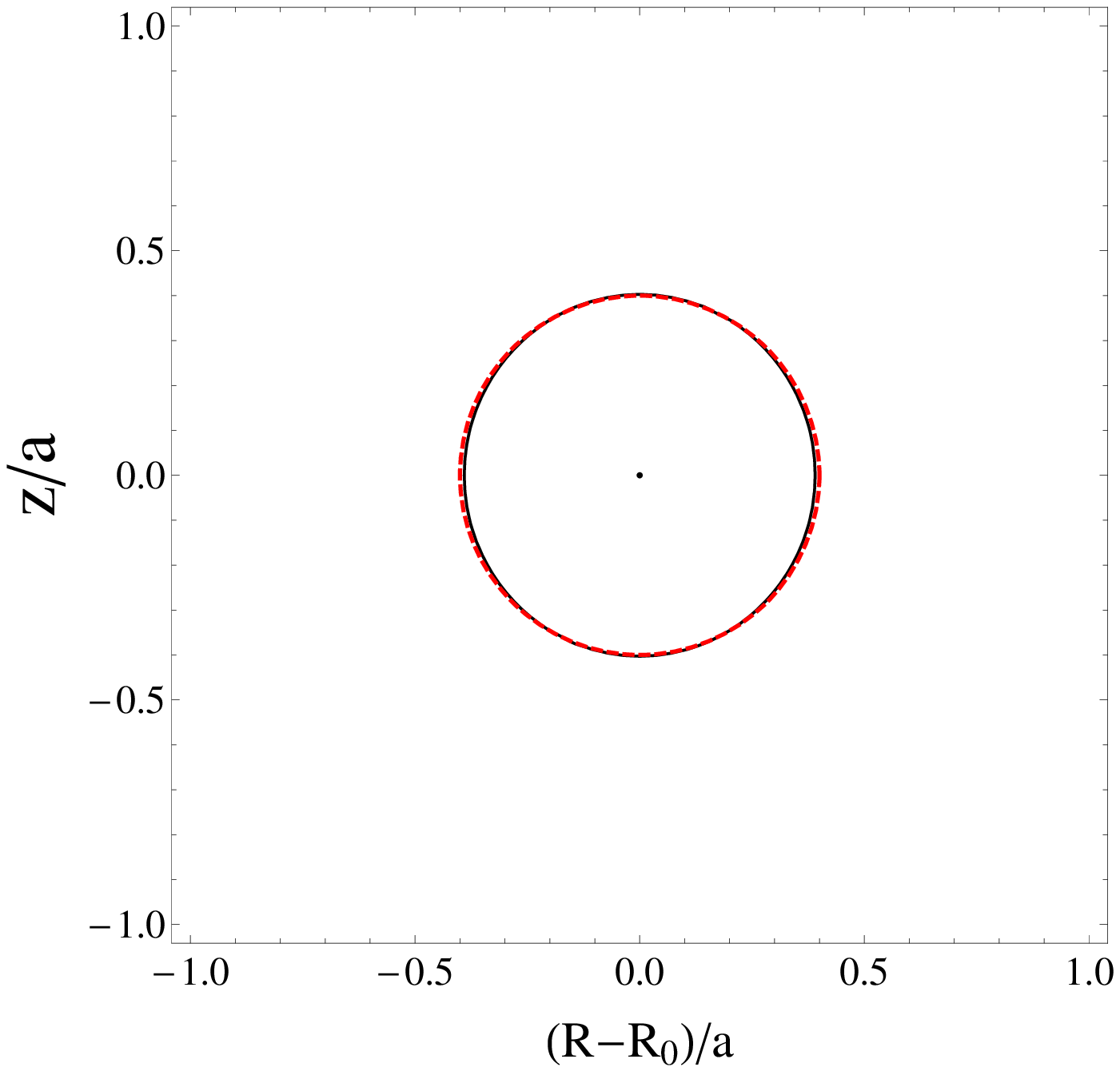}
}
\\
(a)
\\
\parbox{3.in}{
  \includegraphics[scale=0.5]{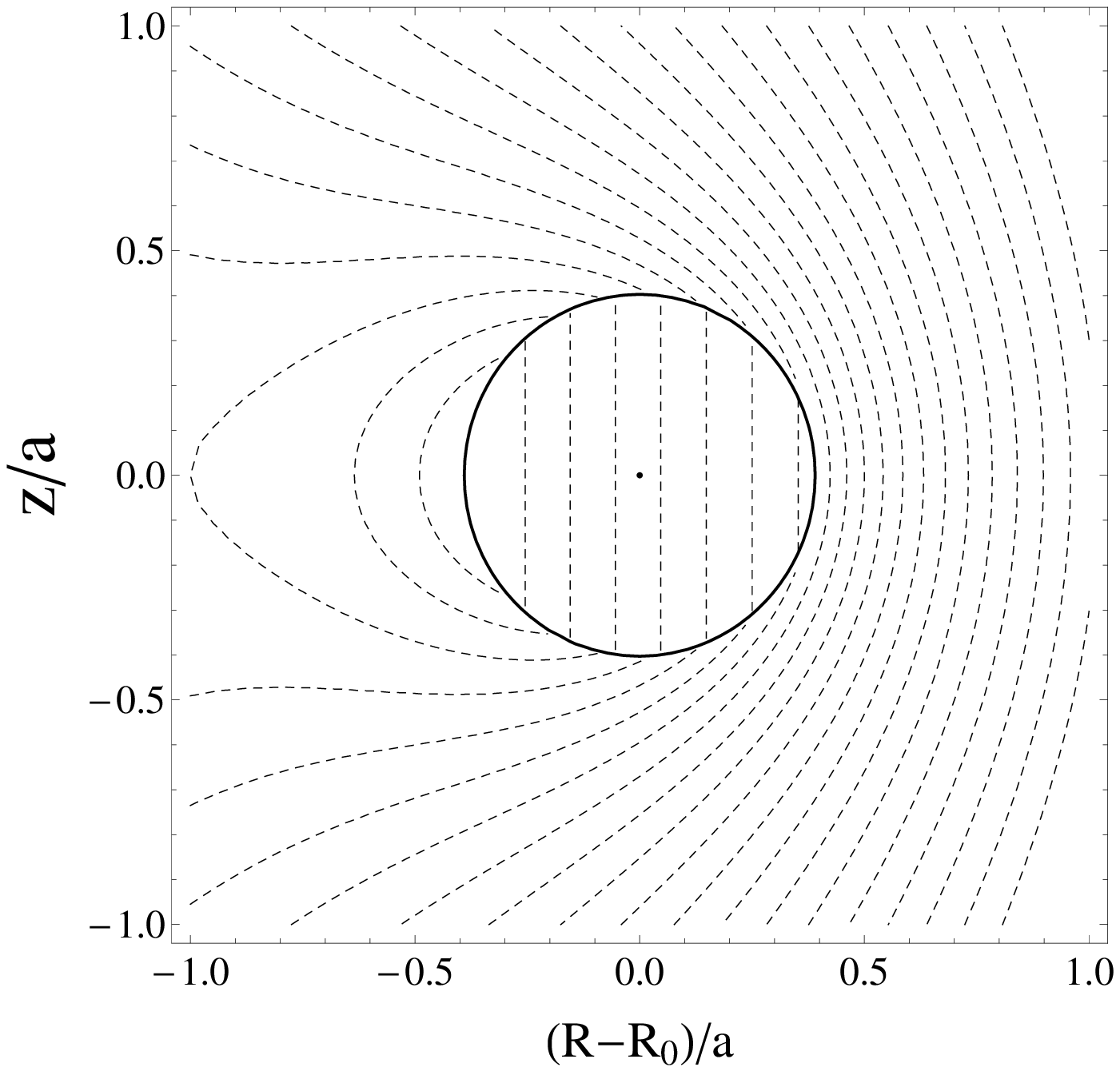}
}
\\
(b)
\etbl
\caption{The runaway electron orbit cross-section in the poloidal
  plane at the beginning of runaway plateau with relativistic factor
  $\gg=100$ and $I_R\sim 0.1$ MA. (a) The comparison between
  runaway orbit with current center at $R_0$ and a circle with minor
  radius $0.4a$. The black solid line represents the runaway orbit, and
  the red dashed line the analytical circle. (b) The runaway orbit in
  the background of total vector potential contour, which is represented
  by black dashed lines. The black dot in both figures denotes the
  position of runaway current center.
}
\label{fig:3}
\end{figure*}

\begin{figure*}
\centering
\noindent
\btbl{c}
\parbox{3.in}{
  \includegraphics[scale=0.5]{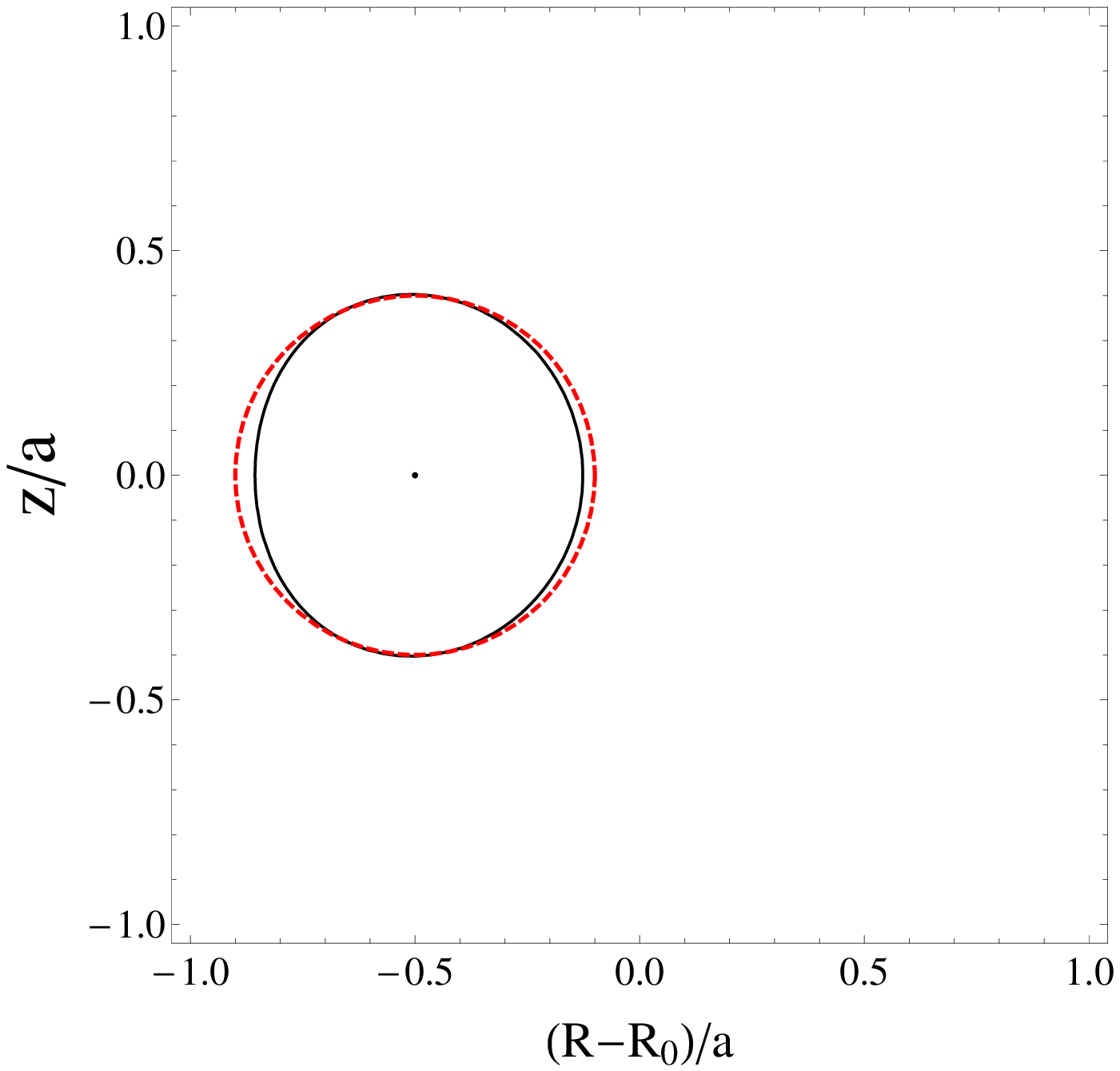}
}
\\
(a)
\\
\parbox{3.in}{
  \includegraphics[scale=0.5]{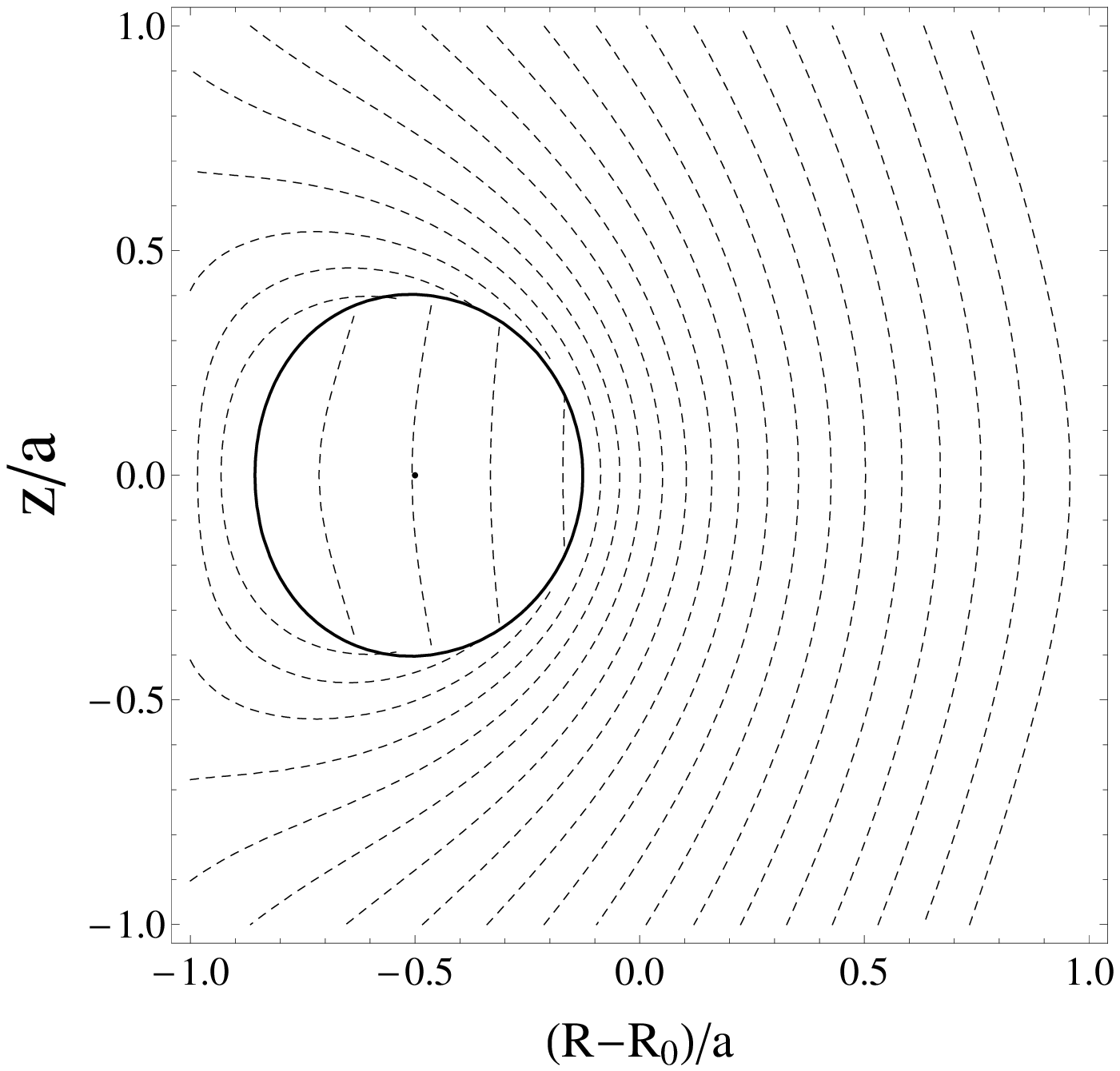}
}
\\
(b)
\etbl
\caption{The runaway electron orbit cross-section in the poloidal
  plane when current center displacement is $d=-0.5a$, corresponding
  $\gg\simeq 68$ with the same $I_R$. (a) The
  comparison between runaway orbit with current center at $R_0$ and a
  circle with minor radius $0.4a$. The black solid line represents the
  runaway orbit, and the red dashed line the analytical circle. (b) The
  runaway orbit in the background of total vector potential contour,
  which is represented by black dashed lines. The black dot in both
  figures denotes the position of runaway current center.
}
\label{fig:4}
\end{figure*}

Due to the symmetry along $\gj$ direction, this system is essentially
2D. It would be convenience for us to express the 2D poloidal plane in
terms of Cartesian coordinates for the purpose of studying runaway
orbit projection in this plane. We choose $x$ to coincide with $R$,
and $y$ to coincide with $Z$. $x=0$ corresponds to $R=R_0$, and $y=0$
corresponds to $Z=0$. Hence the $r$ and $r_i^{(\pm)}$ variables in
Eq.\,(\rfq{eq:RunawayPsi}) and (\rfq{eq:WallPsi}) can be expressed as
\bqbl
\lbq{eq:Cartesian00}
r
=
\sqrt{\left(x-d\right)^2+y^2}
,\eqbl
\bqbl
\lbq{eq:Cartesian10}
r_1^{(+)}
=
\sqrt{\left[x+\left(2a+d\right)\right]^2+y^2}
,\quad
r_2^{(+)}
=
\sqrt{\left(x-d\right)^2+\left(y-4a\right)^2}
\nonumber
,\\
r_3^{(+)}
=
\sqrt{\left[x-\left(2a-d\right)\right]^2+y^2}
,\quad
r_4^{(+)}
=
\sqrt{\left(x-d\right)^2+\left(y+4a\right)^2}
,\eqbl
\bqbl
\lbq{eq:Cartesian20}
r_1^{(-)}
=
\sqrt{\left(x+2a\right)^2+y^2}
,\quad
r_2^{(-)}
=
\sqrt{x^2+\left(y-4a\right)^2}
\nonumber
,\\
r_3^{(-)}
=
\sqrt{\left(x-2a\right)^2+y^2}
,\quad
r_4^{(-)}
=
\sqrt{x^2+\left(y+4a\right)^2}
.\eqbl
Here, $d\equiv R_c-R_0$ is the displacement of runaway current center
relative to the geometric center of the system. Substituting
Eq.\,(\rfq{eq:Cartesian00}) - (\rfq{eq:Cartesian20}) into
Eq.\,(\rfq{eq:RunawayPsi}) and Eq.\,(\rfq{eq:WallPsi}), we then can seek
the constant canonical angular momentum surface for runaways with a
given momentum $p_\|$ by simply solving
Eq.\,(\rfq{eq:AngMomentum}). This surface defines
the runaway orbit in the magnetic field considered in our model. 
In this section, we will consider the displaced runaway orbit for
changing parallel momentum as a sequence of stationary trajectory
surfaces with time dependent terms dropped, each surface corresponds to
a different parallel momentum and a different displacement. Direct
impression of runaway orbit drift with respect to a given change in
parallel momentum can then be obtained by comparing the original runaway
orbit at the beginning of plateau with the decelerated one, as shown in
Fig.\,\ref{fig:3} and Fig.\,\ref{fig:4} respectively. 
Here, the runaway current $I_R$ acts as a given parameter and
does not change in time. The radius of runaway torus is
$a_R=0.4a$. Further, the variation of $p_\|$ due to the inhomogeneity of
magnetic field is negligible as $\gg m_ec^2\muchg\gm B$.

In Fig.\,\ref{fig:3}, the runaway orbit at the beginning of runaway
plateau is shown. The runaway electron parallel momentum is set to be
$p_{\|0}=2e\R{\gm_0I_R}{2\gp}\R{R_0}{a}$, the constant vertical field is chosen
as $B_{Z0}=-p_{\|0}/eR_0$ so that the runaway current center will be at
$R_0$. For runaway current on the order of $I_R\sim 0.1$ MA, the
aforementioned choice of parallel momentum corresponds to a relativistic
factor $\gg=100$. The inverse aspect ratio is chosen as
$\ge=0.2$. The runaway orbit is compared with a analytical 
circle with minor radius being $0.4a$ in Fig.\,\ref{fig:3} (a). In
Fig.\,\ref{fig:3} (b), the orbit is put in the background of 
vector potential contour. The sudden change in the field behavior within the
runaway torus is due to the step functions in Eq.\,(\rfq{eq:RunawayPsi}),
and will not affect the runaway orbit in any way.
Then we consider the case when the runaways have decelerated due to
radiation drag. The relativistic factor is now $\gg\simeq 68$, the
displacement is found by calculating the constant $p_\gj$ contour
iteratively so that the geometric center of orbit matches the current
center position $R_0+d$. The comparison between the orbit and a
analytical circle with minor radius $0.4a$ is also shown in
Fig.\,\ref{fig:4}, as well as the total vector potential contour.

The most important feature obtained from this comparison is that runaway
electrons will drift inward as long as they are decelerating, which will
contribute to the inward runaway current drift observed in runaway
plateau regime. The detailed dynamic of this inward drift will be
discussed in Section \ref{s:OrbitDrift}. Also, it can be seen that the
deviation of runaway transit orbit from circle is less than
$\mathcal{O}\left(\ge\right)$ comparing to $a_R$, justifying 
our assumption that the runaway orbit cross-section can be approximated
as a circle even with substantial displacement. 

\section{Inward drift of runaway electron transit orbit}
\label{s:OrbitDrift}

\vskip1em

The runaway orbit for a given $p_\|$ is demonstrated in Section
\ref{s:AngularMomentum} by iteratively seeking the constant
$p_\gj$ surface. The explicit time dependence of this orbit is dropped.
However, we are also interested in the dynamic of runaway orbit drift
which is more relevant to the control of current displacement. That is,
we wish to know analytically how much the displacement would be for a
given change in runaway momentum $\gD p_\|$. The time scale of this
displacement is also of interest.

This dynamic can be get by considering the energy equation for runaways
along with the modified Euler-Lagrange equation. We write down the
instantaneous change of both energy and angular momentum caused by an
unspecified toroidal force $F$ as follows,
\bqbl
m_ec^2d\gg
=
F\R{p_\|}{m_e\gg}d\gt
,\eqbl
\bqbl
dp_\gj
=
FR d\gt
.\eqbl
Here, we have used the fact that the direction of magnetic field line is
mostly toroidal. Recalling that $p_\|=\gg m_ec$, we have
\bqbl
dp_\|
=
Fd\gt
.\eqbl
Integrating over one bounce period $\gD t$, we have
\bqbl
\gD p_\|
=
\int_0^{\gD t}{Fd\gt}
,\quad
\gD p_\gj
=
\int_0^{\gD t}{FRd\gt}
.\eqbl
While the exact form of $F$ is required to write down the
relationship between $\gD p_\|$ and $\gD p_\gj$,
we will demonstrate that, for particles under effective radiation drag
and externally applied toroidal electric field, we have
\bqbl
\lbq{eq:Qgj}
-e\gD\left(A_{ex}R\right)
+\int_0^{\gD t}{eE_dRd\gt}
=
\gD p_\|\left(R_0+d\right)
.\eqbl
Here, $E_d$ is the effective electric field experienced by the runaway
electrons. 

For externally applied toroidal field, $F$ has the following
form
\bqbl
F
=
eE_{ex0}\R{R_0}{R}
.\eqbl
Integrating along unperturbed transit orbit over one bounce period while
assuming the parallel momentum and field line pitch angle
being near constant within one orbit revolution, the change in momentum
caused by external field is then
\bqbl 
\gD p_\|^{(ex)}
=
eE_{ex0}\R{R_0}{R_0+d}\gD t
+\mathcal{O}\left(\ge^2\right)
.\eqbl
At the same time, according to Eq.\,(\rfq{eq:ExTorField}), the change in
$A_{ex}R$ is
\bqbl
\gD\left(A_{ex}R\right)
=
-E_{ex0}R_0\gD t
.\eqbl
Thus we have
\bqbl
\lbq{eq:Qgj01}
-e\gD\left(A_{ex}R\right)
=
\gD p_\|^{(ex)}\left(R_0+d\right)
.\eqbl
On the other hand, we can write $E_d=E_{sd}+E_{bd}$, where $E_{sd}$ and
$E_{bd}$ represent effective drag field caused by synchrotron and
bremsstrahlung radiation respectively. Assuming $\gg$ and pitch angle
being a near constant across one bounce period, we can write
\cite{Martin1998,Bakhtiari2005}
\bqbl
E_{sd}
=
E_{sd0}\R{R_0^2}{R^2}
,\quad
E_{bd}
=
E_{bd0}
.\eqbl
Hence the parallel momentum change and angular momentum change can be
obtained using similar integration with above to yield
\bqbl
\lbq{eq:DeltaPpar}
\gD p_\|^{(sd)}
=
eE_{sd0}\R{R_0^2}{\left(R_0+d\right)^2}\gD t 
+\mathcal{O}\left(\ge^2\right)
,\quad
\gD p_\|^{(bd)}
=
eE_{bd0}\gD t 
+\mathcal{O}\left(\ge^2\right)
,\eqbl
\bqbl
\gD p_\gj^{(sd)}
=
eE_{sd0}\R{R_0^2}{R_0+d}\gD t
+\mathcal{O}\left(\ge^2\right)
,\quad
\gD p_\gj^{(bd)}
=
eE_{bd0}\left(R_0+d\right)\gD t
+\mathcal{O}\left(\ge^2\right)
.\eqbl
Thus we have
\bqbl
\lbq{eq:Qgj02}
\gD p_\gj^{(d)}
=
\gD p_\gj^{(sd)}
+\gD p_\gj^{(bd)}
=
\left(\gD p_\|^{(sd)}+\gD p_\|^{(bd)}\right)
\left(R_0+d\right)
.\eqbl
Combining Eq.\,(\rfq{eq:Qgj01}) and Eq.\,(\rfq{eq:Qgj02}), we naturally get
Eq.\,(\rfq{eq:Qgj}). 

\begin{figure*}
\centering
\noindent
\btbl{c}
\parbox{3.in}{
  \includegraphics[scale=0.45]{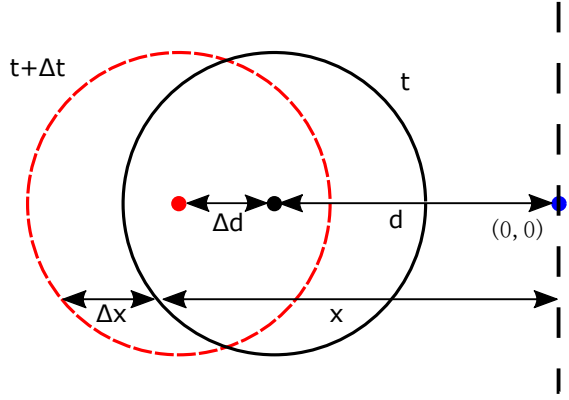}
}
\etbl
\caption{A schematic plot for the runaway transit orbit at time $t$ and
  $t+\gD t$, with current center displacement $d$ and $d+\gD d$
  respectively. The relative major radial position $x$ for a arbitrary
  point on the transit orbit surface, and its displacement $\gD x$ after
  $\gD t$ is also shown on the plot.
}
\label{fig:5}
\end{figure*}

Armed with the knowledge of angular momentum change, we now
proceed to study the drift of runaway orbit. We do this by integrating
Eq.\,(\rfq{eq:ELequationNew}) over $\gD t$ and seek variation of $x$ and
$d$, namely $\gD x$ and $\gD d$, for any given change in parallel
momentum $\gD p_\|$. Due to our assumption of circular cross section, we
have $\gD x\simeq\gD d$ and $\gD y\simeq 0$. A schematic plot for $d$,
$x$, $\gD d$ and $\gD x$ is shown in Fig.\,\ref{fig:5}. Substituting
Eq.\,(\rfq{eq:Qgj}), we write
\bqbl
\lbq{eq:MomentumVar}
e\gD\left[\left(A_R+A_w\right)R\right]
+eB_{Z0}R\gD x
+\gD p_\| \left(x-d\right)
+p_{\|}\gD x
=
0
.\eqbl
Recalling that $B_{Z0}=-p_{\|0}/eR_0$, the above equation is then
rewritten as
\bqbl
\lbq{eq:Essential}
e\gD\left[\left(A_R+A_w\right)R\right]
-\left(p_{\|0}-p_\|\right)\gD x
+\gD p_\| \left(x-d\right)
-p_{\|0}\R{x}{R_0}\gD x
=
0
.\eqbl
Eq.\,(\rfq{eq:Essential}) is the most essential equation in our following
analysis on the horizontal drift of runaway trajectory.

It would be convenient to discuss the two
extreme case where the time scale of runaway displacement being much
longer than the resistive time scale of the wall, and, conversely, the
displacement time scale being much shorter than the resistive time
scale. In the former case, the contribution from wall current vanish,
and the constant vertical magnetic field is crucial in stabilizing the
horizontal drift.  In the latter case, the eddy currents from wall takes
over this role, as their contribution now dominate over that of the
vertical field. Here, we will first study the no wall limit, which is
much simpler than the ideal wall limit. Then we will look into the more
interesting ideal wall case.

\subsection{Runaway drift dynamic with highly resistive wall}
\label{s:NoWallDrift}

\vskip1em

The simpler of the aforementioned two scenarios is the case where the
time scale of current center drift is much longer than the resistive
time scale of the wall. In this case, the wall can be seen as
magnetically transparent. That is, there is no response from wall
current to the change of magnetic field within the vessel.

Under this consideration, $A_w$ vanish from Eq.\,(\rfq{eq:Essential}), and
we have
\bqbl
e\gD\left[\left(A_R+A_w\right)R\right]
=
-e\R{\gm_0I_RR_0}{2\gp}\R{y}{a_R^2}\gD y
.\eqbl
It is important to recognize that the variation of $x$ and $d$ cancel
each other in $\gD\left(A_RR\right)$ since $\gD x\simeq \gD d$.
Also note that although assumed $\gD y$ being small, we still formally
keep it here. We will show that it is indeed small later.
For convenience, we define the following normalized parallel momentum
\bqbl
\lbq{eq:normalization}
\bsp_\|
\equiv
\left(e\R{\gm_0I_R}{2\gp}\R{R_0}{a}\right)^{-1}p_\|
.\eqbl
We further expand the last term at LHS of Eq.\,(\rfq{eq:Essential}) using
$x-d$, so that Eq.\,(\rfq{eq:Essential}) can be written into
\bqbl
\lbq{eq:EssentialNW}
-\R{ay}{a_R^2}\gD y
+\left(\bsp_{\|}-\bsp_{\|0}\right)\gD x
+\gD \bsp_\| \left(x-d\right)
-\bsp_{\|0}\R{d}{R_0}\gD x
-\R{\bsp_{\|0}}{R_0}\left(x-d\right)\gD x
=
0
.\eqbl
Since $x$ can be chosen as any number between $\left[-a_R+d,a_R+d\right]$,
the requirement of $\gD x$ having non-trivial solution for any given
$\gD p_\|$ demands that
\bqbl
\lbq{eq:NWDrift01}
-\R{ay}{a_R^2}\gD y
+\left(\bsp_{\|}-\bsp_{\|0}\right)\gD x
-\bsp_{\|0}\R{d}{R_0}\gD x
=
0
,\eqbl
\bqbl
\lbq{eq:NWDrift02}
\gD x
=
\R{R_0}{\bsp_{\|0}}\gD \bsp_\|
.\eqbl
For consistency, we must also require that
\bqbl
\R{\pa}{\pa d}\left(\bsp_{\|}-\bsp_{\|0}\right)
=
\R{\gD \bsp_\|}{\gD d}
=
\R{\gD \bsp_\|}{\gD x}
.\eqbl
These requirements yield the following solution of equation 
\bqbl
\lbq{eq:NWDrift}
\bsp_\|
-\bsp_{\|0}
=
\bsp_{\|0}\R{d}{R_0}
,\eqbl
\bqbl
\gD y
=
0
.\eqbl
It can be seen that the runaway torus drift exactly in a rigid body
manner. Also, the runaway current drift and the total change in parallel
momentum has a clean and simple linear relation, and the runaway
electrons will drift inward as long as they are decelerating. This
linear behavior derive from the fact that the prescribed vertical field
is the dominant stabilizing term in the drift equation for no wall
limit, rather than the wall current term which is dependent on the
runaway current displacement. It can be further inferred that the
runaways would hit the wall even for some small change in momentum on
the order $\left|p_\|-p_{\|0}\right|\sim\mathcal{O}\left(\ge\right)$.

This result is essentially along the same line with the scenario studied
bu Guan et al. in Ref. \onlinecite{Guan2010}, as both cases concerns the
drift of runaway electrons in a prescribed magnetic field. The only
difference is that Guan et al. studied the outward drift of accelerating
runaways in a constant poloidal field carried by plasma current, while
here we are looking at the inward drift of decelerating runaways in a
constant vertical field sustained by external coils.

It is desirable for us to estimate the time scale of aforementioned
horizontal drift caused by effective radiation drag. This can be done by
combining Eq.\,(\rfq{eq:DeltaPpar}) and Eq.\,(\rfq{eq:NWDrift02}), and
estimating the no wall limit drift time scale as $\gt_{nw}=a/\left(\gD
x/\gD t\right)$. For our case considered here, $I_R=0.1$MA, $\ge=0.2$,
so that $\R{\gm_0I_RR_0}{2\gp a}= 1\times10^{-1}V\cdot s/m$,
The effective drag field can be estimated by considering synchrotron
radiation and bremsstrahlung radiation \cite{Martin1998,Bakhtiari2005},
with $\gg_0\sim 100$, $B_{T0}\sim 3$T, and $R_0=5$m. The
resulting effective drag field is on the order of $1.19$ V/m. Hence the
characteristic time scale of runaway orbit  drift is $\gt_d\sim
3.4\times 10^{-2}$s. 

\subsection{Runaway drift dynamic with ideally conducting wall}
\label{s:IdealWallDrift}

\vskip1em

Now, we proceed to consider the case where the time scale of runaway
displacement is much shorter than the resistive time of the wall. Thus
the wall can be seen as ideally conducting as studied in Section
\ref{s:AngularMomentum}. The algebra is more complicated than that of
Section \ref{s:NoWallDrift} due to the complicated nature of $A_w$, but
the method is along the same line.

To simplify the expression, we now formally write
$\gD\left[\left(A_R+A_w\right)R\right]$ in term of
$\gL^{(i)}\left(x,d,y\right)\gD x+M^{(i)}\left(x,d,y\right)\gD y$, where
$\gL^{(i)}$ and $M^{(i)}$ have the dimension of inverse length. So
that
\bqbl
e\gD\left(A_RR\right)
=
-e\R{\gm_0I_RR_0}{2\gp}
\left[
\gL\gD x
+M\gD y
\right]
,\eqbl
\bqbl
e\gD\left(A_w^{(+)}R\right)
=
e\R{\gm_0I_RR_0}{2\gp}
\left[
\gL^{(+)}\gD x
+M^{(+)}\gD y
\right]
,\eqbl
\bqbl
e\gD\left(A_w^{(-)}R\right)
=
-e\R{\gm_0I_RR_0}{2\gp}
\left[
\gL^{(-)}\gD x
+M^{(-)}\gD y
\right]
.\eqbl 
The detailed expression for each term is then as follows.

Once again, the contribution from runaway current itself is
\bqbl
\lbq{eq:complicated00}
e\gD\left(A_RR\right)
=
-e\R{\gm_0I_RR_0}{2\gp}\R{\gD r}{a_R}
=
-e\R{\gm_0I_RR_0}{2\gp}\R{y}{a_R^2}\gD y
.\eqbl
Meanwhile, the contribution from eddy current is
\bqbl
e\gD\left(A_w^{(\pm)}R\right)
=
\pm e\R{\gm_0I_RR_0}{2\gp}
\left[
\R{\gD r_1^{(\pm)}}{r_{1}^{(\pm)}}
+\R{\gD r_2^{(\pm)}}{r_{2}^{(\pm)}}
+\R{\gD r_3^{(\pm)}}{r_{3}^{(\pm)}}
+\R{\gD r_4^{(\pm)}}{r_{4}^{(\pm)}}
\right]
.\eqbl
Here, for $A_w^{(-)}$, we have
\bqbl
\lbq{eq:complicated10}
\R{\gD r_1^{(-)}}{r_{1}^{(-)}}
=
\R{\left(x+2a\right)}
{\left(x+2a\right)^2
+y^2}\gD x
+\R{y}
{\left(x+2a\right)^2
+y^2}\gD y
,\eqbl
\bqbl
\lbq{eq:complicated11}
\R{\gD r_2^{(-)}}{r_2^{(-)}}
=
\R{x}
{x^2
+\left(y-4a\right)^2}\gD x
+\R{\left(y-4a\right)}
{x^2
+\left(y-4a\right)^2}\gD y
,\eqbl
\bqbl
\lbq{eq:complicated12}
\R{\gD r_3^{(-)}}{r_3^{(-)}}
=
\R{\left(x-2a\right)}
{\left(x-2a\right)^2
+y^2}\gD x
+\R{y}
{\left(x-2a\right)^2
+y^2}\gD y
,\eqbl
\bqbl
\lbq{eq:complicated13}
\R{\gD r_4^{(-)}}{r_4^{(-)}}
=
\R{x}
{x^2
+\left(y+4a\right)^2}\gD x
+\R{\left(y+4a\right)}
{x^2
+\left(y+4a\right)^2}\gD y
.\eqbl
On the other hand, for $A_w^{(+)}$, we have
\bqbl
\lbq{eq:complicated20}
\R{\gD r_1^{(+)}}{r_{1}^{(+)}}
=
\R{2\left[x-d+2\left(a+d\right)\right]}
{\left[\left(x-d\right)+2\left(a+d\right)\right]^2
+y^2}\gD x
+\R{y}
{\left[\left(x-d\right)+2\left(a+d\right)\right]^2
+y^2}\gD y
,\eqbl
\bqbl
\lbq{eq:complicated21}
\R{\gD r_2^{(+)}}{r_{2}^{(+)}}
=
\R{\left(y-4a\right)}
{\left(x-d\right)^2
+\left(y-4a\right)^2}\gD y
,\eqbl
\bqbl
\lbq{eq:complicated22}
\R{\gD r_3^{(+)}}{r_{3}^{(+)}}
=
\R{2\left[x-d-2\left(a-d\right)\right]}
{\left[\left(x-d\right)-2\left(a-d\right)\right]^2
+y^2}\gD x
+\R{y}
{\left[\left(x-d\right)-2\left(a-d\right)\right]^2
+y^2}\gD y
,\eqbl
\bqbl
\lbq{eq:complicated23}
\R{\gD r_4^{(+)}}{r_{4}^{(+)}}
=
\R{\left(y+4a\right)}
{\left(x-d\right)^2
+\left(y+4a\right)^2}\gD y
.\eqbl
It is found that $M^{(+)}-M^{(-)}$ is two order of magnitude smaller
than $M$, and will be omitted hereafter. Hence we can finally write
down the following form,
\bqbl
e\gD\left[\left(A_R+A_w\right)R\right]
=
e\R{\gm_0I_RR_0}{2\gp}
\left[
\left(
\gL^{(+)}
-\gL^{(-)}
\right)\gD x
-M\gD y
\right]
.\eqbl
Also, result from Section \ref{s:AngularMomentum} indicate that in ideal
wall limit $p_\|-p_{\|0}$ being comparable with $p_{\|0}$, hence the
last term at the LHS of Eq.\,(\rfq{eq:Essential}) is now next order
effect, and can be neglected.

\begin{figure*}
\centering
\noindent
\btbl{c}
\parbox{3.in}{
  \includegraphics[scale=0.45]{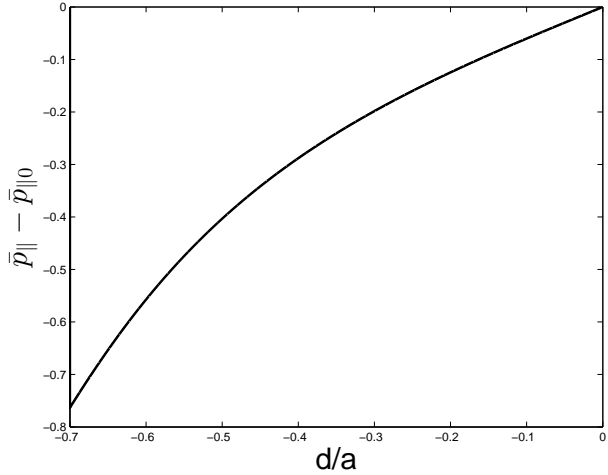}
}
\etbl
\caption{The relation between runaway current center displacement $d$
  and the total change in normalized parallel momentum
  $\bsp_\|-\bsp_{\|0}$. It is assumed that $a=1$ and $a_R=0.3$.
}
\label{fig:6}
\end{figure*}

Using the same normalization Eq.\,(\rfq{eq:normalization}),
Eq.\,(\rfq{eq:Essential}) in ideal wall limit can then be written as
\bqbl
a\left(
\gL^{(+)}
-\gL^{(-)}
\right)\gD x
-aM\gD y
-\left(\bsp_{\|0}-\bsp_\|\right)\gD x
+\gD\bsp_\| \left(x-d\right)
=
0
.\eqbl
Expanding $\gL^{(\pm)}$ using $x-d$, we have
\bqbl
\gL^{(+)}\left(x,d\right)
=
\gL^{(+)}\big|_{x=d}
+\left(\pa_x\gL^{(+)}\right)\big|_{x=d}\left(x-d\right)
+\mathcal{O}\left(\left(x-d\right)^2\right)
,\\
\gL^{(-)}\left(x,d\right)
=
\gL^{(-)}\big|_{x=d}
+\left(\pa_x\gL^{(-)}\right)\big|_{x=d}\left(x-d\right)
+\mathcal{O}\left(\left(x-d\right)^2\right)
.\eqbl
Once again, the requirement of non-trivial solution demands that
\bqbl
\lbq{eq:ICom00}
a\left(
\gL^{(+)}\big|_{x=d}
-\gL^{(-)}\big|_{x=d}
\right)\gD x
-aM\big|_{x=d}\gD y
-\left(\bsp_{\|0}-\bsp_\|\right)\gD x
=
0
,\eqbl
\bqbl
\lbq{eq:ICom01}
\gD x
=
-\R{x-d}{a}
\left[
\left(\gL^{(+)}-\gL^{(+)}\big|_{x=d}\right)
-\left(\gL^{(-)}-\gL^{(-)}\big|_{x=d}\right)
\right]^{-1}
\gD\bsp_\|
.\eqbl
Also, for consistency, we require that,
\bqbl
\lbq{eq:ICom02}
\R{\pa}{\pa d}\left(\bsp_{\|}-\bsp_{\|0}\right)
=
\R{\gD \bsp_\|}{\gD d}
=
\R{\gD \bsp_\|}{\gD x}
.\eqbl
Taking the limit $x\rightarrow d$,
Eq.\,(\rfq{eq:ICom01}) then can be solved numerically by simple 4th order
Runge-Kutta method \cite{ButcherBook2008} to obtain the
displacement of runaway torus regarding to any parallel momentum
change. The result of numerical integration is shown in
Fig.\,\ref{fig:6} for parameters $a=1$ and $a_R=0.3$. Again, it can be
seen that the runaway electrons will drift inward as long as they are
losing momentum, regardless of the detailed history of deceleration.

\begin{figure*}
\centering
\noindent
\btbl{c}
\parbox{5in}{
  \includegraphics[scale=0.55]{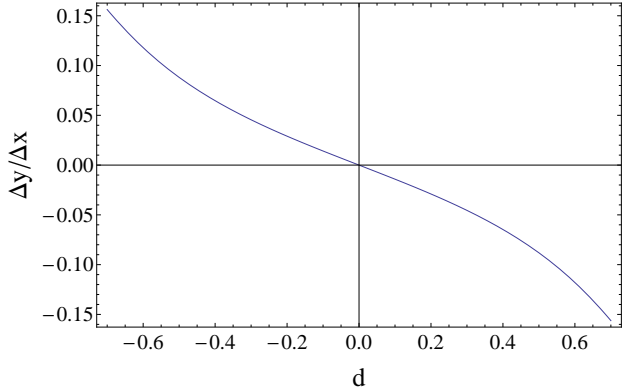}
}
\etbl
\caption{The ratio between $\gD y$ and $\gD x$ as a function of
  displacement $d$, calculated assuming
  $a=1$ and $a_R=0.3$. It can be seen that $\gD y$ is indeed much
  smaller than $\gD x$ even for substantial displacement of runaway
  torus, confirming the validity of our previous assumption that $\gD
  y\simeq 0$.
}
\label{fig:7}
\end{figure*}

It is noteworthy that the orbit displacement described by
Eq.\,(\rfq{eq:ICom01}) is not that of a rigid body, as opposed to the no
wall limit case. Indeed, it can be seen from Eq.\,(\rfq{eq:ICom00}) and
Eq.\,(\rfq{eq:ICom01}) that $\gD x$ is actually dependent on $x-d$, and
$\gD y$ is not exactly zero. This corresponds to the ``squeeze'' of
runaway torus cross-section seen in Fig.\,\ref{fig:4}. Thus it is
desirable for us to check the magnitude of $\gD y$ and $\gD x-\gD d$ for
the consistency of our model. For this purpose, it's possible to write
down analytical solutions for $\gD y$. Checking through
Eq.\,(\rfq{eq:complicated10}) - (\rfq{eq:complicated23}), it can be
shown that 
\bqbl
\pa_d\left(\gL^{(+)}\big|_{x=d}\right)
=
2\left(\pa_x\gL^{(+)}\right)\big|_{x=d}
,\eqbl
\bqbl
\pa_d\left(\gL^{(-)}\big|_{x=d}\right)
=
\left(\pa_x\gL^{(-)}\right)\big|_{x=d}
.\eqbl
Hence we can infer from Eq.\,(\rfq{eq:ICom00}) - (\rfq{eq:ICom02}) that
$\bsp_\|$ has the following relation with the displacement,
\bqbl
\bsp_\|-\bsp_{\|0}
=
\left(
\gL^{(-)}\big|_{x=d}
-\R{1}{2}\gL^{(+)}\big|_{x=d}
\right)a
.\eqbl
Meanwhile, we also have
\bqbl
\gD y
\lbq{eq:small}
=
\R{1}{2}\left(M\big|_{x=d}\right)^{-1}\left(\gL^{(+)}\right)\big|_{x=d}\gD x
.\eqbl
The ratio $\gD y/\gD x$ can then be calculated using Eq.\,(\rfq{eq:small})
for given $a$ and $a_R$. Consider $a=1$, $a_R=0.3$ and $y$ being
positive as an example, $\gD y$ turns out to be indeed much smaller than
$\gD x$ even for substantial displacement, as can be seen in
Fig.\,\ref{fig:7}. Hence the vertical displacement can indeed be
neglected as an next order effect, and our assumption that $\gD y\simeq
0$ stands valid.

\begin{figure*}
\centering
\noindent
\btbl{c}
\parbox{5in}{
  \includegraphics[scale=0.55]{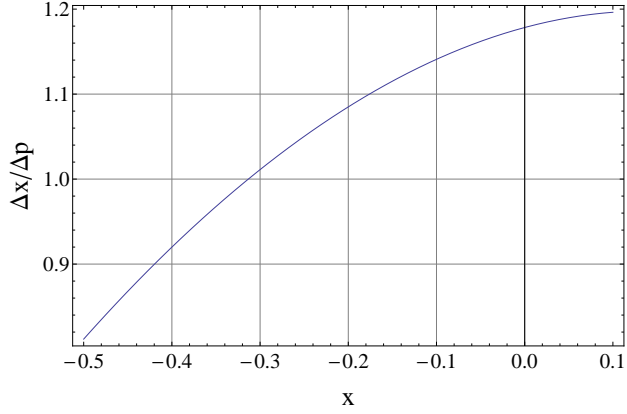}
}
\etbl
\caption{The drift rate $\gD x/\gD\bsp_\|$ as a function of $x$ for
  $a=1$, $a_R=0.3$ and $d=-0.2$. At $x=d$, we have $\gD x=\gD d$. It can
  be seen that the trajectory displacement can be divided into a
  dominant rigid body displacement and a secondary deformation. 
}
\label{fig:8}
\end{figure*}

\begin{figure*}
\centering
\noindent
\btbl{c}
\parbox{5in}{
  \includegraphics[scale=0.45]{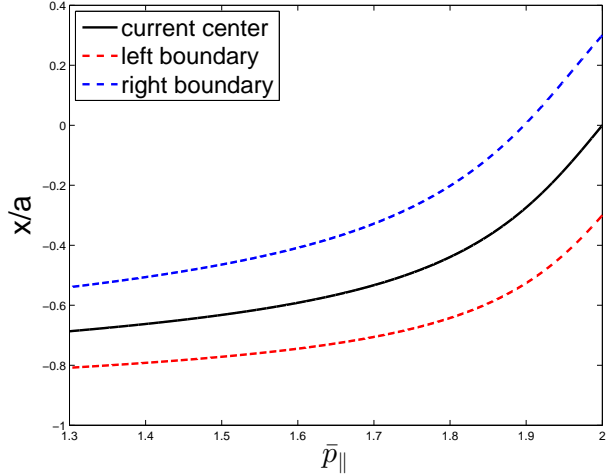}
}
\etbl
\caption{The position of current center and both extreme points of the
  runaway torus as functions of the current center displacement. It can
  be seen that there is only minimal deformation of the circular even
  for significant displacement of current center.
}
\label{fig:9}
\end{figure*}

Meanwhile, the deviation of $\gD x$ from $\gD d$ can be obtained by
considering Eq.\,(\rfq{eq:ICom01}) for given $a$, $a_R$ and $d$. As an
example, we choose $a=1$, $a_R=0.3$ and $d=-0.2$, the corresponding
drift rate $\gD x/\gD \bsp$ as a function of coordinate $x$ is shown in
Fig.\,\ref{fig:8}. It can be seen that the trajectory evolution can be
separated into a dominant rigid body displacement and a secondary
deformation which tend to ``squeeze'' the runaway torus as it drift
towards the wall. To demonstrate the consistency of our model, we
numerically integrate Eq.\,(\rfq{eq:ICom01}) to show that such
deformation actually have minimal impact on the over all shape of
runaway cross-section until the runaways come really
close to the wall. Assuming the same initial runaway torus radius and a
initial normalized parallel momentum $\bsp_{\|0}=2$ (which corresponds
to $\gg\sim 100$ in our case), the
displacement of both the left and right extreme points of the runaway
torus, as well as that of the current center is shown in
Fig.\,\ref{fig:9}. It is apparent that the deformation of torus
cross-section only becomes important when the current center is rather
close to the wall, 

We can also estimate the time scale of horizontal drift in ideal wall
limit by considering the deceleration caused by radiation drag. Once
again, we have, $\R{\gm_0I_RR_0}{2\gp a}= 1\times10^{-1}V\cdot s/m$, and
the effective radiation drag can still be estimated as on the order of
$1.19$ V/m. At the same time, when the runaway torus is not so close to
the war, the drift rate $\gD
x/\gD\bsp_\|\sim\mathcal{O}\left(1\right)$, hence the characteristic
time scale of horizontal drift is $\gt_d\sim 8.4\times 10^{-2}$s. This
time scale reasonably agree with experimental observation, where the
current center moves one third of the minor radius in $25$ms
\cite{Zhang2012}. This corresponds to a time scale about $8.75\times
10^{-2}$s. It should be noted that the estimation here is made by using
the radiative drag experienced by $\gg\sim 100$ runaways, as the runaway
decelerate, the drift velocity is expected to be slower, hence the
actual time for runaways to hit the wall may be somewhat longer than
estimated here.

\section{Discussion and Conclusion}
\label{s:Conclusion}

\vskip1em

The inward drift of runaway current center during runaway plateau is
studied in this paper. This horizontal drift motion is required by the
balance between change in canonical angular momentum and the mechanical
angular momentum change caused by radiation drag. We are mainly
interested in the plateau regime after disruption where most of the
current is carried by runaway electrons themselves. In this
consideration, for any drift of runaway electron relative to the field
line, the current center itself will also drift. Since the magnetic
field lines is generated by this runaway current, the resulting current
center drift motion is essentially non-linear, as opposed to the linear
drift motion of test particle runaways studied in previous works
\cite{Guan2010}. 

The runaway transit orbit surface is obtained by seeking the constant
canonical angular momentum surface in a unperturbed 2D equilibrium. It
is found that runaways will always drift inward as long as they are
losing momentum. The eddy current and external vertical field are found
to play a crucial role in stabilizing this horizontal drift, without
which the runaways will not stop until they hit the first wall even for
small amount of momentum loss. The dynamic of this inward drift is
analyzed by taking the variation of canonical angular momentum and
electron energy, which yield a first order ODE describing the trajectory
displacement for any given change in parallel momentum. The remarkable
feature of this drift motion is that is does not really depends on the
detailed history of deceleration, only on how much momentum is lost in
total. The time scale of such displacement is estimated by using models
of effective radiation drag. The time scale thus calculated reasonably
agrees with experimental observation. 

It is noteworthy that the horizontal drift we discussed here has
drastically different physics with the force imbalance along major
radius, which has been invoked when discussing the observed inward
motion during plateau regime \cite{Eidietis2012}. The
fundamental physics here is the balance in canonical angular momentum
budget, which can not be recovered by simply considering the runaway
current as an ordinary current carrying circuit. An easy way to see this
is by considering a runaway torus in perfect force balance. We then
consider a certain loss of parallel momentum, with minimal decrease in
the velocity of runaways. The change in $\waJ_\gj\times\waB_Z$ force
balance is negligible, so that if we only consider the runaway current
as an ordinary circuit with finite mass, the previously force balanced
current will still be in equilibrium along major radial direction, thus
it should not move at all. However, as we have seen in Lagrangian
mechanics analysis, the runaways will actually drift inward in
response to the change in mechanical angular momentum. Hence our study
here provided a new powerful mechanism which may play an important role
in analyzing runaway motions during plateau regime.

Strong simplification has been made to ensure the runaway current drift
we concerned here to be analytically tractable. In a more realistic
consideration, various more complicated model such as finite
distribution of runaways in phase space and the impact of finite
resistive wall should be included. Most importantly, the runaway beam
with finite spatial distribution along minor radius could be of great
interest, as the interaction between different ``rings'' of runaway
torus may produce more complicated picture than that is studied
here. Nonetheless, our simplified model has captured the most basic and
fundamental trend for runaway trajectory behavior, namely the inward
drift trend for decelerating runaways. Tracking the evolution of
aforementioned more complicated model require numerical tools, and it is
left for future works. 

\vskip1em
\centerline{\bf Acknowledgments}
\vskip1em

  The authors thank C. Liu, X.-G. Wang and A. Bhattacharjee for
  fruitful discussion. The authors also thank an anonymous referee
  for constructive comments. This work is partially supported by
  National Magnetic Confinement Fusion Energy Research Project under
  Grant No.\,2015GB111003, National Natural Science Foundation of China
  under Grant No.\,1126114032, 11575185, 11575186 and 11305171,
  JSPS-NRF-NSFC A3 Foresight Program under Grant No.\,11261140328,
  the China Scholarship Council and US DoE
  contract No.\,AC02-09-CH11466. 

 
\end{document}